\documentstyle[preprint,prb,aps]{revtex}
\begin{document}
\draft
\title{Histogram Monte Carlo study of next-nearest-neighbor
 Ising antiferromagnet on a stacked triangular lattice}
\author{M.L. Plumer, A. Mailhot, R. Ducharme, and A. Caill\'e}
\address{ Centre de Recherche en Physique du Solide et D\'epartement de
Physique}
\address{Universit\'e de Sherbrooke, Sherbrooke, Qu\'ebec, Canada J1K 2R1}
\author{H.T. Diep}
\address{ Groupe de Physique Statistique, Universit\'e de Cergy-Pontoise}
\address{ 47-49, Avenue des Genottes, B.P. 8428, 95806 Cergy-Pontoise
Cedex, France }
\date{January 1993}
\maketitle
\begin{abstract}
Critical properties of the Ising model on a stacked triangular
lattice, with antiferromagnetic first and second-neighbor
in-plane interactions, are studied by extensive histogram Monte Carlo
simulations.  The results, in conjunction with the recently determined
phase diagram, strongly suggest that the transition from the
period-3 ordered state to the paramagnetic phase remains in the xy
universality class.  This conclusion is in contrast with a previous
suggestion of mean-field tricritical behavior.
\end{abstract}

\section{Introduction}

There is little consensus in recent literature regarding critical
phenomena associated with the simple stacked triangular antiferromagnet
\cite{plumA}.  In the cases of Heisenberg and xy spin models, noncolinear
magnetic ordering gives rise to nontrivial symmetry considerations and
this has lead to contrasting proposals on the nature of the temperature-driven
phase transition from the period-3 state to the paramagnetic phase: The
suggestion by Kawamura
\cite{kawaA} of new universality classes is not included in the scenario
of nonuniversality put forth by Azaria {\it et al.},\cite{aza} where
first-order, mean-field tricritical or O(4) criticality can occur depending
details of the model (also see Refs. [\onlinecite{chub,kawaB,saul,kunz}]).
 Even the Ising model on this frustrated lattice has a highly degenerate
ground state \cite{wan,cop} and unusual temperature and magnetic-field
induced phase transitions.\cite{netz}  The conclusion from symmetry
arguments is that the paramagnetic transition belongs to the standard xy
universality class,\cite{blank} consistent with preliminary Monte Carlo
results.\cite{mats}  More recently, this simple picture has been challenged
by Heinonen and Petschek \cite{hein} (hereafter referred to as HP) who made
the remarkable proposal
that this transition is mean-field tricritical, as inspired by their
Monte Carlo analysis of critical exponents, the structure factor, and the
observation that a sufficiently large value of third-neighbor in-plane
interaction $J_3$ induces a strong first-order transition to a different
type of order.  (Such an idea has also found support from recent
experimental results.\cite{fark})  It thus appears
that, independently, and for completely different reasons, the suggestion
of tricriticality associated with the stacked triangular antiferromagnet
was made by both HP (for the Ising model) and by
Azaria {\it et al.} (for the Heisenberg model).  It is the purpose of
this work to re-examine the scenario proposed by HP through
means of more extensive (and more accurate) conventional and histogram
Monte Carlo simulations of the Ising model with antiferromagnetic
first and second-neighbor in-plane interactions ($J_1,J_2>0$).

A number of studies have been made of the effects of further-neighbor
interactions on the triangular antiferromagnet.
\cite{kats,cafl,joli}  At a critical value $J_2>0$, the period-3
state (C3) is destabilized in favor of magnetic order with a periodicity of
two (C2).
Of particular interest to the present work are the results of recent
Monte Carlo determinations of $J_2-T$ (temperature) phase diagrams for
Ising \cite{nag} as well as xy and Heisenberg \cite{lois} models.  The
paramagnetic transition temperature $T_N$ decreases sharply with increasing
$J_2$ until the critical value, after which $T_N$ increases sharply.  With
$J_1=1$, the
critical values of $J_2$ were found to be approximately 0.10 for the Ising
model and 0.125 for xy and Heisenberg models.  For each model, the transition
line to the C3 state was found to be continuous, within the accuracy of
these conventional simulations, whereas the C2 transition is always first
order.  In the case of Heisenberg and xy models, a continuous transition
to an incommensurate (IC) order is observed at larger values of $J_2$ (e.g.,
$J_2\approx 1.1$ in the Heisenberg case).

A simple mean-field argument explains the principal features of these phase
diagrams, and also reveals a connection with the results of HP.
For systems governed by the Hamiltonian
\begin{equation}
{\cal H}~=~\frac12
\sum_{ij} J_{ij} {\bf S}_i \cdot {\bf S}_j ,
\label{hamil}
\end{equation}
the Fourier transform of the exchange interaction J({\bf Q}) determines
the wavevector which characterizes the spin
modulation of the first ordered state to stabilize as the temperature is
lowered.  (This result is independent of the number of spin components.)
In the case of ferromagnetic interactions along the $c$ axis ($J_0<0$),
$Q_\|=0$ and the modulation occurs entirely in the basal plane.
We consider here $J({\bf Q}_\bot)$ with up to third-neighbor in-plane
interactions included. \cite{plumB}
Wavevectors which maximize this function give the
desired result, shown in Fig. 1.  With $J_3=0$, the C2 phase
is stabilized for $\frac18 \leq J_2 \leq 1$ and the IC state is realized for
$J_2\geq 1$.  As found by HP, the C2 state also occurs
with $J_2=0$ and a ferromagnetic third-neighbor interaction.  Their critical
value $J_3\sim-0.08$ can be compared with the present mean-field result
$J_3=-\frac19 \simeq -0.111$.
The scenario put forth by these authors calls for interactions between
primary C3 and secondary C2 fluctuations to drive the transition first
order (to the C3 phase) at very small values $J_3<0$.  Thus $J_3\simeq0$ would
be a tricritical point (not to be confused with the multicritical point where
C3, C2 and paramagnetic phases meet).  In view of the phase diagram Fig. 1
and the results of Ref.[\onlinecite{nag}], such behavior should also be
revealed
by considering the effects of small $J_2$ (as done in the present
study).  Finally, note that the dramatic depression in $T_N$
at the multicritical point observed in the Monte Carlo results of
Ref.[\onlinecite{nag}] is seen from Fig. 1 to likely be a result of large
critical fluctuations since the IC and C2 phases are degenerate at this
value of $J_2$.  The relatively weak dependence of $T_N$ on $J_3<0$
observed by HP is consistent with this explanation.

Although the conventional Monte Carlo simulations of Ref.[\onlinecite{nag}]
suggest that the transition with
$0 \leq J_2 < 0.1$ remains continuous, the
implementation of more sensitive finite-size scaling techniques may be
required to detect a very weak first order transition, as might be
expected near a tricritical point.  The Ferrenberg-Swendsen histogram
Monte Carlo method \cite{ferrA,fuk,ferrB} has proven useful for
this purpose, especially when used to determine the limiting value
of the (internal) energy cumulant \cite{chal}
\begin{equation}
U(T)~=~1 - \frac13<E^4>/<E^2>^2.
\label{cumulant}
\end{equation}
This quantity has a minimum at a phase transition, with the property
$U(T_N) \rightarrow U^* = \frac23$ in the infinite-lattice limit.
The histogram
method allows for the possibility of precise determination of extrema
exhibited by other thermodynamic functions at $T_N$.  The scaling
behavior with system size of these quantities can yield accurate
estimates of critical exponents in the case of a continuous transition,
or reveal simple volume dependence if the transition is first order.
The utility of this type of Monte Carlo method for the present
purposes is very nicely described by Reimers {\it et al.} \cite{reim}
Recently, Bunker {\it et al.} \cite{bunk} made such histogram
analysis of the present model in the case of nearest-neighbor in-plane
interactions only.  They find critical exponents consistent with
xy universality, in contrast with HP.  Our work corroborates and extends
their results.

Guided by the results of Ref.[\onlinecite{nag}], we test the proposals
of HP by performing Monte Carlo simulations at three values of
$J_2$: 0, 0.08, and 0.25.  Conventional Monte Carlo analysis at $J_2=0$
reveals that the critical region is rather narrow in
temperature, thus providing an expalanation for the erroneous
exponent estimates given by HP.  Extensive histogram simulations
at $J_2=0.08$ provide convincing evidence that the transition remains
continuous and of xy universality.  The results at $J_2=0.25$ serve
as an example of scaling behavior in the case of a first order
transition to the C2 state.

\section{$J_2=0$}

Conventional Monte Carlo simulations were performed for the
case of nearest-neighbor in-plane
interactions only.  Runs of 2x$10^4$ to $10^5$ Monte Carlo steps (MCS)
per spin were made with the initial 4x$10^3$ to 2x$10^4$ MCS discarded
for thermalization.  In the case of smaller runs, quantites were averaged over
4 independent simulations using random initial spin configurations.  Periodic
boundary conditions on lattices LxLxL with L=12-30 were used.
Finite-size scaling of only the critical exponent $\beta$ was considered, using
the C3 order parameter $M \sim t^\beta$ [$t=(T_N-T)/T_N)$] defined in terms of
a Fourier component as in Ref.[\onlinecite{plumC}].
The extrapolation technique of Landau \cite{land} was
used on results close to the transition temperature, known from HP to
be near 2.9$J_1$, to estimate their values for $L \rightarrow \infty$.
The results are presented in Fig. 2, where L=12 data were excluded in the fit
for the highest four temperatures.
$T_N$ was then adjusted to yield the best linear fit of the extrapolated
data for $ln(M_\infty)$ {\it vs.} $ln(t)$ plots.

Initial analysis performed on data for $12 \leq L \leq 24$ and
$2.0 \leq T \leq 2.7$ yielded values $T_N \simeq 2.82$ and
$\beta \simeq 0.19$, close to those of HP.  However, a more detailed study
using data at larger L and higher T revealed a different set of results.
It was difficult to achieve good linear fits of all the data between
T=2.00 and T=2.90 so that some of the lower-temperature points were excluded.
The outcome of this procedure was instructive.
Using only the six highest temperatures (T=2.60-2.90) gave the estimates
$T_N \simeq 2.92$ and $\beta \simeq 0.30$.
Fig. 3 shows results with only the four highest temperatures included,
which yield the estimates
$T_N \simeq 2.93(1)$ and $\beta \simeq 0.33(3)$.
The latter values are dramatically different from those of HP but
are consistent with more accurate histogram results (see below
and Ref.[\onlinecite{bunk}]) and the expected xy universality.
The above analysis strongly suggests that this model exhibits an
unusually narrow critical region of temperature.  This may be a
consequence of the proximity of the C2-order instability and the coupling
of these fluctuations to the primary C3 order, as suggested to be
important for this model by HP.

Such a system is ideally suited to exhibit the power of the histogram
method, where exponents can be extracted from simulations performed {\it very}
close to the transition temperature.  Since these simulations were done
mainly as a test for later
runs at $J_2=0.08$, and to corroborate the more detailed study
of Ref.[\onlinecite{bunk}], only a single histogram was made for each
lattice size L=12,15,18,21,24,30 at the temperature T=2.93 with $1.2$x$10^6$
MCS per spin, where $2$x$10^5$ MCS were discarded for thermalization.
More details are presented in the next section for the
case of $J_2=0.08$ and, in the interest of brevity, we present no data here
but state only the results of our analysis.
Finite-size scaling of the maxima (or minima) for the specific heat ($C$),
susceptibility ($\chi$), energy cumulant ($U$), and logarithmic derivative
of the order parameter \cite{ferrB}
\begin{equation}
V(T)~=~<ME>/<M> - <E>,
\label{logM}
\end{equation}
yield estimates for $\alpha / \nu$, $\gamma / \nu$, $U^*$, and $1/ \nu$,
respectively.  In addition, $M(T_N)$ scales as $L^{- \beta / \nu}$,
where $T_N$ can be adjusted to give the best
linear fit.  The results $\alpha / \nu < 0.2(3)$,
$\beta / \nu = 0.50(3)$ (using $T_N=2.928$), $\gamma / \nu = 2.03(6)$,
$1 / \nu = 1.46(5)$, and $U^* \simeq 0.666667(5)$ were obtained.
(Errors are difficult to assign; those given here are estimated from
the robustness of the linear fitting.  Evaluation of results
from many simulations are necessary to obtain more reliable estimates
of statistical errors).
These values compare favorably with renormalization-group results
$\alpha / \nu = -0.018$, $\beta / \nu = 0.519$, $\gamma / \nu = 1.96$,
$1 / \nu = 1.49$.
The quantity $(U^* - U)$ scales as $L^{-m}$, where in principle
$m = \alpha / \nu$.
However, only very weak maxima in $C$ and minima in $U$ were observed,
\cite{chal} consistent with a very small value of $\alpha$.  As noted in
Ref.[\onlinecite{chal}], in such cases $(U^* - U)$ may exhibit simple volume
dependence.  The observed exponent $m \simeq 2.75$ is consistent with
this expectation.

\section{$J_2=0.08$}

Finite-scaling analysis of histogram Monte Carlo simulations for the
case of $J_2=0.08$ are presented here.  A relatively large value of
second-neighbor coupling was desired to enhance the possibility of
observing first-order effects, but not too close to the multicritical
point at $J_2 \simeq 0.1$ where larger fluctuations might be expected.
However, since $J_2$ is quite small relative to the other energy
scale in the model ($J_1=1$), longer runs are necessary to fully
realize its effects (and achieve the same accuracy as with $J_2=0$).
With these considerations, it can be expected that the extraction of critical
exponents will be less reliable than for the case of $J_2=0$.

Simulations were performed using the same parameters as described above
for the case $J_2=0$, but at two or three different temperatures near the
transition estimated from the results of Ref.[\onlinecite{nag}] to be
$T_N \simeq 2.2$.
This ensured that the histogram was made at a temperature close enough to
the extrema of the thermodynamic quantity of interest for the extraction
of meaningful results.  As a general guide to reliable data, the temperature
of the extrema should lie within the range determined by half of the maximum
of the histogram.  If it does not, a new histogram should be generated at
a temperature closer to the extrema of the desired function.

Logarithmic scaling plots were made of the various thermodynamic
functions as described in the previous section.  Convincing evidence
that the transition remains continuous is found in results for the
energy cumulant, where the value $U^*$=0.666655(20) was extracted.
Fig. 4 displays results for $(U^*-U)$, with the assumtion $U^*=\frac23$;
the good linear fit further supports this conclusion.
The specific heat (as well as $U$)
again exhibited only very weak temperature maxima, with a small dependence
on L, as displayed in Fig. 5.  It can be concluded from these data
that $\alpha / \nu$ is very small.  Scaling of the other functions produced
the estimates $\beta / \nu = 0.50(3)$ (using $T_N=2.197$),
$\gamma / \nu = 2.07(6)$, $1 / \nu = 1.40(5)$, and $m=3.10(10)$, obtained
from all of the data.  Using data from only the three largest lattice sizes
yielded slightly different results for $\gamma / \nu = 2.00(6)$,
$1 / \nu = 1.42(5)$, and $m=3.04(10)$.  In an effort to demonstrate further
that our data are consistent with xy universality, the scaling plots
shown in Figs. 6-9 were made using the renormalization-group exponents.
The results are convincing on this point.  We note
that no visual discrimination could be made between these plots and
those made with our extracted exponents.

\section{$J_2=0.25$}

Finally, a brief summary of histogram results for the case of
$J_2=0.25$ where a relatively strong first-order transition to the C2 phase
is expected.\cite{nag}  The main purpose
of these simulations is to illustrate that the energy cumulant is
sensitive to the order of the transition for the present model.
Since the ordered state has a periodicity of two, even values of the
lattice size L were used.
Simulations were perfomed with L=12,18,24,30 at a range of temperatures
guided by considerations as outlined in the previous section,
near the known transition point, $T_N \simeq 2.66$.
Very large fluctuations were found
in the results for the $\chi$ and $V$, likely a consequence of metastability
effects.  Finite-size scaling of these data was not fruitful.
Fig. 10 displays scaling results for the minima in $U$, which were
used to obtain the
saturation value $U^*$=0.6595(3).  This result, along with a reasonably
good linear fit of the specific heat maxima $vs.$ $L^3$, confirms the
strong first-order nature of this transition and further strengthens the
conclusions of the previous section.

\section{Summary and Conclusions}

The results presented in this work and in Refs.[\onlinecite{nag} and
\onlinecite{bunk}] strongly suggest that the phase transition
associated with the period-3 state of the next-nearest-neighbor
Ising antiferromagnet on a stacked triangular lattice exhibits
xy universality.   In contrast with the proposal by Heinonen and Petschek,
no mean-field tricritical behavior is observed.  The idea by these authors
that the coupling of fluctuations to the nearby period-2 order should be
important, however, appears to be responsible for a shrinking of the true
critical region and offers an explanation for their erroneous estimates
of critical exponents.  The histogram Monte Carlo method was found
to be ideally suited for the exposition of the real critical behavior
in this system.  It remains to be tested if these conclusions
are relevant in the case of continuous spin models on this frustrated
lattice where there are diverse proposals regarding the criticality.

\acknowledgements
We thank A. Bunker, A. Chubukov, A. Ferrenberg, and O. Heinonen for
useful discussions and D. Loison for technical assistance.  M.L.P.
is grateful to the Universit\'e de Cergy-Pontoise for the hospitality
which promoted this study.  This work was also supported by NSERC of
Canada and FCAR du Qu\'ebec.
%

\begin{figure}
\caption{Phase diagram determined by maximizing $J({\bf Q}_\bot)$ with up
to third-neighbor in-plane interactions (with antiferromagnetic
first-neighbor coupling $J_1=1$) where C2 and C3 represent commensurate phases
of periodicity two and three, respectively, and IC denotes the
incommensurate phase.  Solid and broken lines indicate first and
second-order transitions, respectively.}
\label{fig1}
\end{figure}

\begin{figure}
\caption{Finite-size scaling of the order parameter at $J_2=0$ and selected
temperatures T=2.00-2.90 for lattice sizes L=12-30.
L=12 values were excluded from the fit for the four highest
temperatures.}
\label{fig2}
\end{figure}

\begin{figure}
\caption{Ln-ln plot of the order-parameter values for $J_2=0$ extrapolated to
$L \rightarrow \infty$ from the results of Fig. 2.  Only the
four highest temperatures were used in the linear fit, with the
results $T_N \simeq 2.93(1)$ and $\beta \simeq 0.33(3)$.}
\label{fig3}
\end{figure}

\begin{figure}
\caption{Scaling of the energy-cumulant minima for $J_2=0.08$ with $U^*$ set
to $\frac23$.  The resulting estimate for $m$ is given by the slope.}
\label{fig4}
\end{figure}

\begin{figure}
\caption{Temperature behavior of the specific heat showing the
weak maxima and small lattice-size dependence for L=12 (lower curve)
to L=30 (upper curve).}
\label{fig5}
\end{figure}

\begin{figure}
\caption{Scaling behavior of the specific heat maxima assuming
the renormalization-group value of $\alpha / \nu$ for the xy
universality class.}
\label{fig6}
\end{figure}

\begin{figure}
\caption{Scaling behavior of the order parameter at $T_N=2.195$ assuming
the renormalization-group value of $\beta / \nu$ for the xy
universality class.}
\label{fig7}
\end{figure}

\begin{figure}
\caption{Scaling behavior of the susceptibility maxima assuming
the renormalization-group value of $\gamma / \nu$ for the xy
universality class.}
\label{fig8}
\end{figure}

\begin{figure}
\caption{Scaling behavior of the maxima in the logarithmic derivative
of the order parameter (3) assuming the renormalization-group value
of $1 / \nu$ for the xy universality class.}
\label{fig9}
\end{figure}

\begin{figure}
\caption{Scaling of the energy-cumulant minima with L
for the case $J_2=0.25$.  The extrapolated estimate for
$U^*$ is 0.6595(3).}
\label{fig10}
\end{figure}

\end{document}